\RequirePackage[ l2tabu , orthodox ]{nag}
\documentclass[ a4paper , reprint , aps , prb ]{revtex4-1}
\usepackage{scrextend}
\usepackage{mathtools}
\usepackage[dvipsnames,pdftex,hyperref]{xcolor} 
\usepackage{graphicx}
\usepackage[ caption = false ]{subfig}
\usepackage{siunitx}
\usepackage[ language = british ]{chemmacros} 
\usepackage[pdftex,unicode,colorlinks,linkcolor=RedViolet,citecolor=BlueViolet,filecolor=magenta,urlcolor=Mahogany]{hyperref}

\begin{document}
\title{Defects and oxidation resilience in InSe}
\author{K.J.~Xiao}
  \email{c2dxk@nus.edu.sg}
  \affiliation{Centre for Advanced 2D Materials and Graphene Research Centre, National University of Singapore, Singapore 117542, Singapore}
\author{A.~Carvalho}
  \email{carvalho@nus.edu.sg}
  \affiliation{Centre for Advanced 2D Materials and Graphene Research Centre, National University of Singapore, Singapore 117542, Singapore}
\author{A.~H.~Castro~Neto}
  \affiliation{Centre for Advanced 2D Materials and Graphene Research Centre, National University of Singapore, Singapore 117542, Singapore}
  \affiliation{Boston University, 590 Commonwealth Avenue, Boston, Massachusetts 02215, USA}
\date{\today}
\begin{abstract} \noindent We use density functional theory to study
  intrinsic defects and oxygen related defects in indium selenide.  We
  find that \ch{InSe} is prone to oxidation, but however not reacting
  with oxygen as strongly as phosphorene.  The dominant intrinsic
  defects in \ch{In}-rich material are the \ch{In} interstitial, a
  shallow donor, and the \ch{Se} vacancy, which introduces deep
  traps. The latter can be passivated by oxygen, which is
  isoelectronic with \ch{Se}. The dominant intrinsic defects in
  \ch{Se}-rich material have comparatively higher formation energies.
\end{abstract}

\maketitle
\section{Introduction}
Amongst two-dimensional materials, the families of chalcogenides such
as transition metal dichalcogenides, group-III and IV
monochalcogenides often offer the advantages of stability and the
possibility of fabrication by epitaxial growth methods that can be
scaled up---such as vapor transport epitaxy of chemical vapor
deposition (CVD),\cite{yu-RSC-6-6705} and chemical vapor
transport.\cite{InSe-Ching-Hwa-Ho-Thickness-dependent-carrier-transport}
Indium selenide,\cite{nanoResearchInSe2014} which shares the same
crystal structure with \ch{GaS} and
\ch{InS},\cite{zolyomi2014electrons} has recently been mechanically
exfoliated into few layer
flakes.\cite{deckoff2016observing,mudd2013tuning,bandurin2016high}
Thin \ch{InSe} flakes have been used for phase change memory devices
and image
sensing,\cite{lei-nl503505f,gibson2005phase,robertsonUsefulInSe} and
has been suggested to be a functional material for water
splitting.\cite{H2fromH20-InSe} With respect to the electronic
properties, few layer \ch{InSe} has been shown to have an
extraordinary electron mobility exceeding \num{E3} and
\SI[per-mode=symbol]{E4}{\raiseto{-2}\centi\metre\per\volt\per\second}
at room and liquid-helium temperatures, in few layers, making it one
of the highest known mobility 2D
materials.\cite{bandurin2016high,AbInitioElMobInSe} This is consistent
with the bulk electron mobility, which is also the highest amongst
isomorphic group-III chalcogenides, according to Hall effect
measurements.\cite{segura1984electron} Even though it is often \( n
\)-type, \ch{InSe} can also be \( p \)-type and in that case it can be
interesting for different purposes: It has a very high effective mass
for holes near the \( \Gamma \) point, where there is a
`Mexican-hat'-type van-Hove
singularity.\cite{DFT-tight-binding,parabolic2RSVBM,zolyomi2014electrons,Mudd2016}
Such a singularity gives rise to a ferromagnetic instability at low
temperatures.\cite{multiferroic2D} Different from other materials with
`Mexican-hat'-type bands such as \ch{SnO}, the singularity is present
in the valence band both for monolayer and for few-layer
material.\cite{Mudd2016}

Thus, since both \( p \)- and \( n \)-type conduction regimes are of
technological interest, it is desirable to be able to effectively
control the type and amount of defects and impurities unintentionally
introduced.  \ch{Sn} and \ch{Pb}, when present, can act respectively
as a shallow donor and shallow acceptor.  The first is often cited as
the origin of the \( p \)-type conductivity.  However, intrinsic
shallow donors that cannot be ascribed to any impurity and disappear
upon annealing have been found as well.\cite{segura1984electron,
  segura1983investigation, martinez1992shallow} These were speculated
to be related to \ch{Se} deficiency.\cite{martinez1992shallow}
According to previous theoretical calculations, adsorbed or
interstitial \ch{In} has low formation energy in \ch{In}-rich
material,\cite{robertsonUsefulInSe} parallel to what has been found
for the \ch{Ga} interstitial in \ch{GaS},\cite{Chen2015} However, many studies of
point defects in III-VI materials have been restricted to vacancies or
substitutional type
defects.\cite{Rak2008,Rak2009,Li2017,robertsonUsefulInSe,H2fromH20-InSe,Chen2015elB}
Thus, specific defect signatures of the intrinsic shallow donors have
not been assigned yet.

Interstitial atoms are supposed to increase the mechanical hardness of
bulk \ch{GaSe} by coupling the planar
layers,\cite{kokh2011growth,huang2017experimental} and the same has
been found for other ionized dopants as well.\cite{rak2010doping}

In addition to intrinsic defects, it is important to investigate the
defects caused by the interaction with oxygen and other atmospheric
contaminants.  The recently achieved high mobility transistor devices
were fabricated with \ch{BN}-encapsulated \ch{InSe} layers, that were
thus prevented from contact with the
atmosphere.\cite{bandurin2016high} Still, \ch{InSe} seems to be
relatively stable in contact with air, as cleaved bulk surfaces show
no signs of degradation at room
temperature,\cite{myake-JJAP-23-172,balakrishnan2017engineering}
comparing e.g.~with phosphorene.

In this article, we will provide a detailed theoretical account of the
properties of intrinsic defects and oxygen-related defects in
\ch{InSe}. In addition, we will discuss their impact on the electronic
properties of the material, in particular discussing the identity of
the shallow donors in unintentionally doped \ch{InSe}.

\section{Methods}
\subsection*{Parameters}
The first principles calculations were performed by the density
functional theory (DFT)\cite{dft-hk, dft-ks} implementation known as
{\scshape Quantum ESPRESSO}.\cite{QE-2009}\( ^{,} \)\footnote{version
  6} All of the computations were done consistently using the
following parameters. The pseudopotentials used were given by the
projector augmented wave (PAW)\cite{paw-blochl, paw-from-us}
approximation, and the exchange-correlation functional chosen was the
generalized gradient approximation parametrized by Perdew, Burke, and
Ernzerhof (GGA-PBE).\cite{pbe-gga-made-simple} Specifically, the
PSeudopotential Library (PSL)\cite{DalCorso2014337}\( ^{,}
\)\footnote{versions 0.3.1 and 1.0.0} were used. A plane wave basis
with kinetic energy cutoff of \SI{42}{Ry} was used, and the \( k
\)-point samples in the Brillouin zone were calculated with the \(
\Gamma \)-centered \( 4 \times 4 \times 1 \)
Monkhorst-Pack\cite{monkhorst-pack} grid unless otherwise
specified. Defect ionization transition levels were calculated with a
\( k \)-point grid of \( 8 \times 8 \times 1 \) centered upon \(
\Gamma \), with relaxation. All transition levels presented were at
most \SI{.02}{\electronvolt} from their values when calculated with
the smaller \( k \)-point grid.  All geometries were relaxed to at
least the default convergence thresholds (Forces \( < \SI{E-3}{a.u.}
\)). The vacuum spacing along the \( z \)-axis was six times the
lattice parameter of the primitive cell of the pristine monolayer, to
avoid spurious interactions. All supercells consisted of \( 3 \times 3
\) primitive unit cells.

Finally, to find the migration activation energies for the relevant
defects, we also performed nudged elastic band calculations, without
climbing images nor spins. 


\subsection*{Formation Energies \& Transition Levels}
The formation energy of defect \( D \) is given by
\begin{equation}
  E_{\!f} ( D ) = E ( D ) - \sum_i n_i \mu_i
\end{equation}
where \( E ( D ) \) is the energy of the supercell containing the
defect, and \( n_i \) and \( \mu_i \) are the number of atoms of
species \( i \) and its chemical potential, respectively.  The
chemical potentials were evaluated both in the \ch{In}-rich and
\ch{Se}-rich limit. In the \ch{In}-rich case, the \ch{In} potential
was obtained from the elemental material in the \(\alpha\)-\ch{In},
tetragonal form.  The \ch{Se} chemical potential \(
\mu_{\ch{Se},\text{\ch{In}-rich}} \) in the \ch{In}-rich regime was
obtained from the constraint
\begin{equation}
 E(\text{PS}) = \sum_{j} n_j \mu_{j,\text{\ch{In}-rich}}.
\end{equation}
where PS is the pristine supercell. A similar definition was used to
obtain the chemical potentials in the \ch{Se}-rich limit for which we
used the trigonal \textit{hP3} \ch{Se} allotrope as reference. The
chemical potential for oxygen is obtained from molecular oxygen.

The defect ionization transition levels \( E_D ( q / q + 1 ) \),
defined by the Fermi level at which the formation energy of the
defects in charge state \( q \) is the same as in charge state \( q +
1 \), were found using the marker method, which is more accurate for
2D systems due to the cancellation of systematic
errors\cite{defectsIn2Dvs3D}.  The ionization potential \( I_D \) and
electron affinity \( A_D \) are defined by
\begin{align}
  I_D &= E ( D^+ ) - E ( D^0 ), & A_D &= E ( D^0 ) - E ( D^- ).
\end{align}
The transition levels for acceptors \( E_D ( - / 0 ) \) (donors \( E_D
( 0 / + ) \)) relative to valence band maximum \( E_v \) (downwards
from conduction band minimum \( E_c \)), are given by
\begin{subequations} \label{eq:def:E_D}
\begin{align}
  E_D ( - / 0 ) - E_v &= E_g - \left [ E_c - E_D ( - / 0 ) \right ] =
  E_g - \left [  A_D - A_{PS} \right ] \\
  E_c - E_D ( 0 / + ) &= E_g - \left [ E_D ( 0 / + ) - E_v \right ] =
  E_g - \left [  I_{PS} - I_D \right ]
\end{align}
\end{subequations}
\section{Results}
\subsection{Intrinsic Point Defects}
\begin{figure*}
\includegraphics[resolution=300,width=\textwidth]{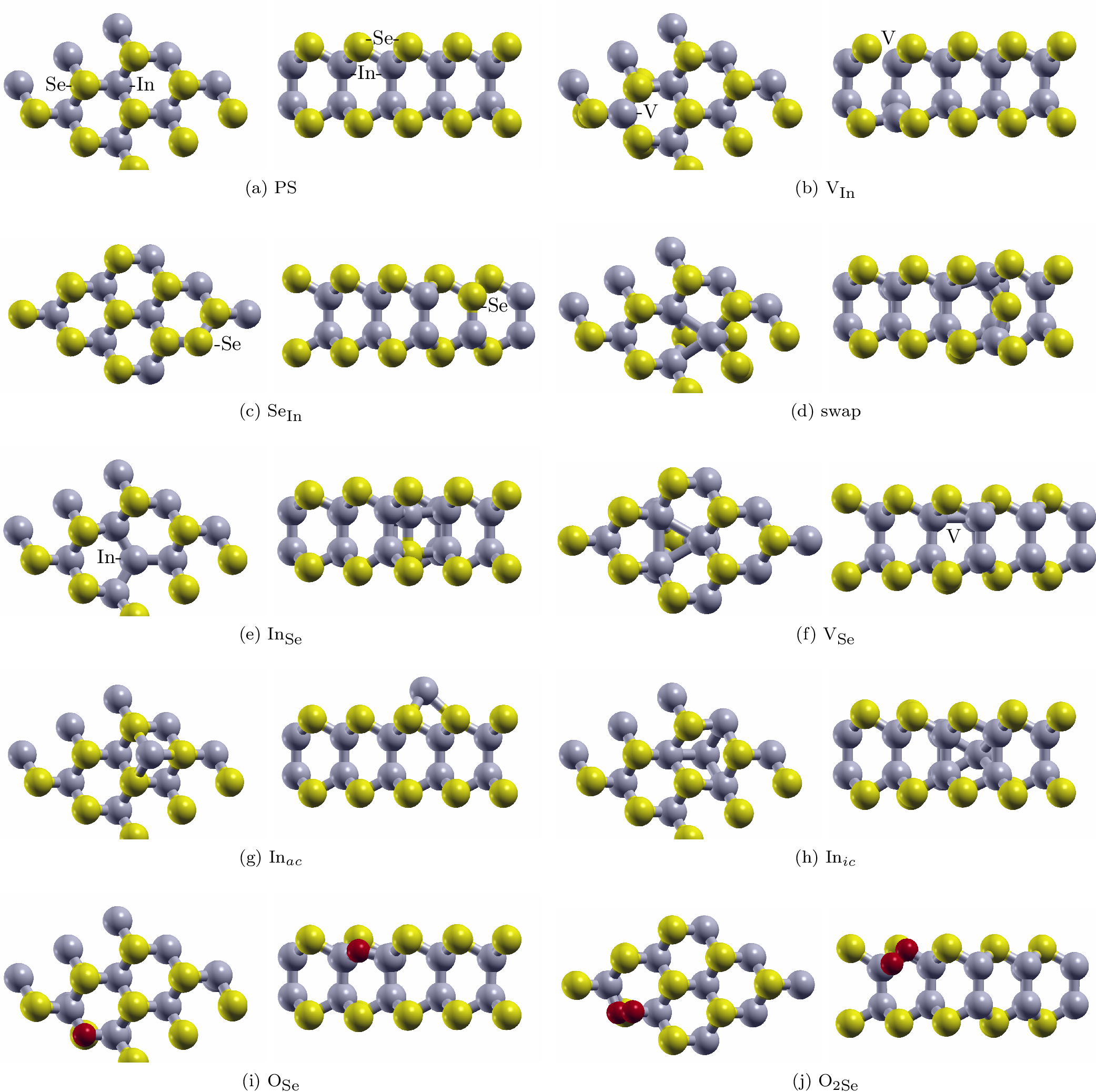}
\caption{\label{fig:ptDxcrysden}(Color online) Top (0001) and side
  (11\(\bar{2}\)0) views of various intrinsic point defects and
  substitutional oxygen in monolayer \ch{InSe}, grouped by
  similarity. 
  (a) PS: pristine supercell. 
  (b) V\(_{\ch{In}}\): indium vacancy. 
  (c) \ch{Se}\(_{\ch{In}}\): selenium-in-indium anti-site. 
  (d) swap: swapping adjacent selenium and indium. 
  (e) \ch{In}\(_{\ch{Se}}\): indium-in-selenium anti-site. 
  (f) V\(_{\ch{Se}}\): selenium vacancy. 
  (g) \ch{In}\(_{ac}\): indium hovering above the center of the
  hexagonal interstitial cage. 
  (h) \ch{In}\(_{ic}\): interstitial indium at center of hexagonal
  cage. 
  (i) \ch{O}\(_{\ch{Se}}\): oxygen atom substituting a selenium. 
  (j) \ch{O2}\(_{\ch{Se}}\): oxygen molecule substituting a selenium.}
\end{figure*}

\begin{figure}
\includegraphics[resolution=300,width=\columnwidth]{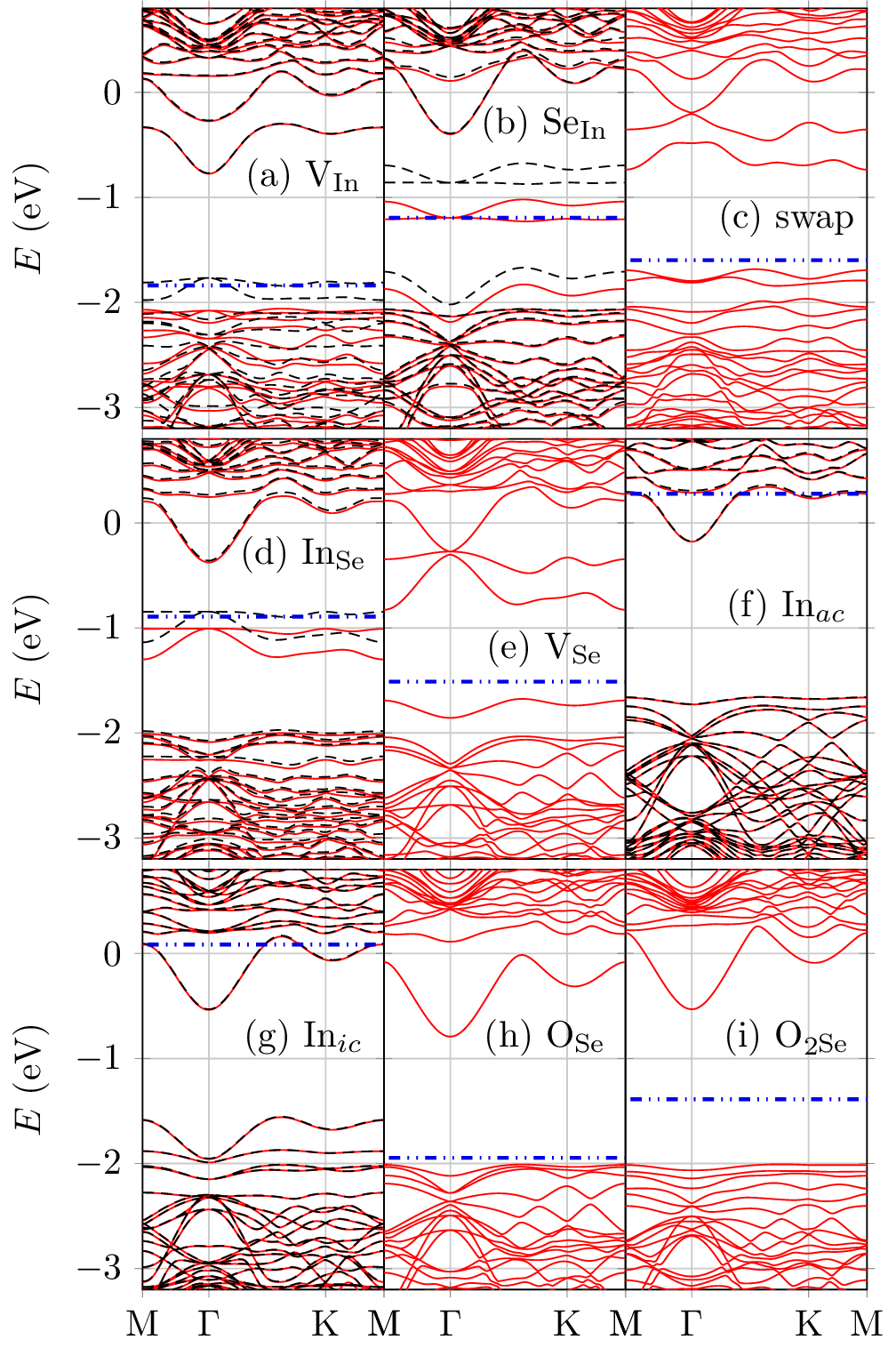}
\caption{\label{fig:bs:intrinsic}(Color online) DFT band structure
  plots of various intrinsic point defects and substitutional oxygen
  defects in monolayer \ch{InSe}: 
  (a) V\(_{\ch{In}}\): indium vacancy.
  (b) \ch{Se}\(_{\ch{In}}\): selenium-in-indium anti-site.
  (c) swap: swapping adjacent selenium and indium.
  (d) \ch{In}\(_{\ch{Se}}\): indium-in-selenium anti-site.
  (e) V\(_{\ch{Se}}\): selenium vacancy.
  (f) \ch{In}\(_{ac}\): indium hovering above the center of the
  hexagonal interstitial cage.
  (g) \ch{In}\(_{ic}\): interstitial indium at center of hexagonal
  cage.
  (h) \ch{O}\(_{\ch{Se}}\): oxygen atom substituting a selenium.
  (i) \ch{O2}\(_{\ch{Se}}\): oxygen molecule substituting a selenium.
   Refer to Fig.~\ref{fig:ptDxcrysden}
  for the respective defects. 
  Majority and minority spin bands are represented by
  continuous and dashed lines, respectively.
  Fermi levels are represented by blue dash-dotted horizontal lines. }
\end{figure}

\begin{figure}
\includegraphics[resolution=300,width=\columnwidth]{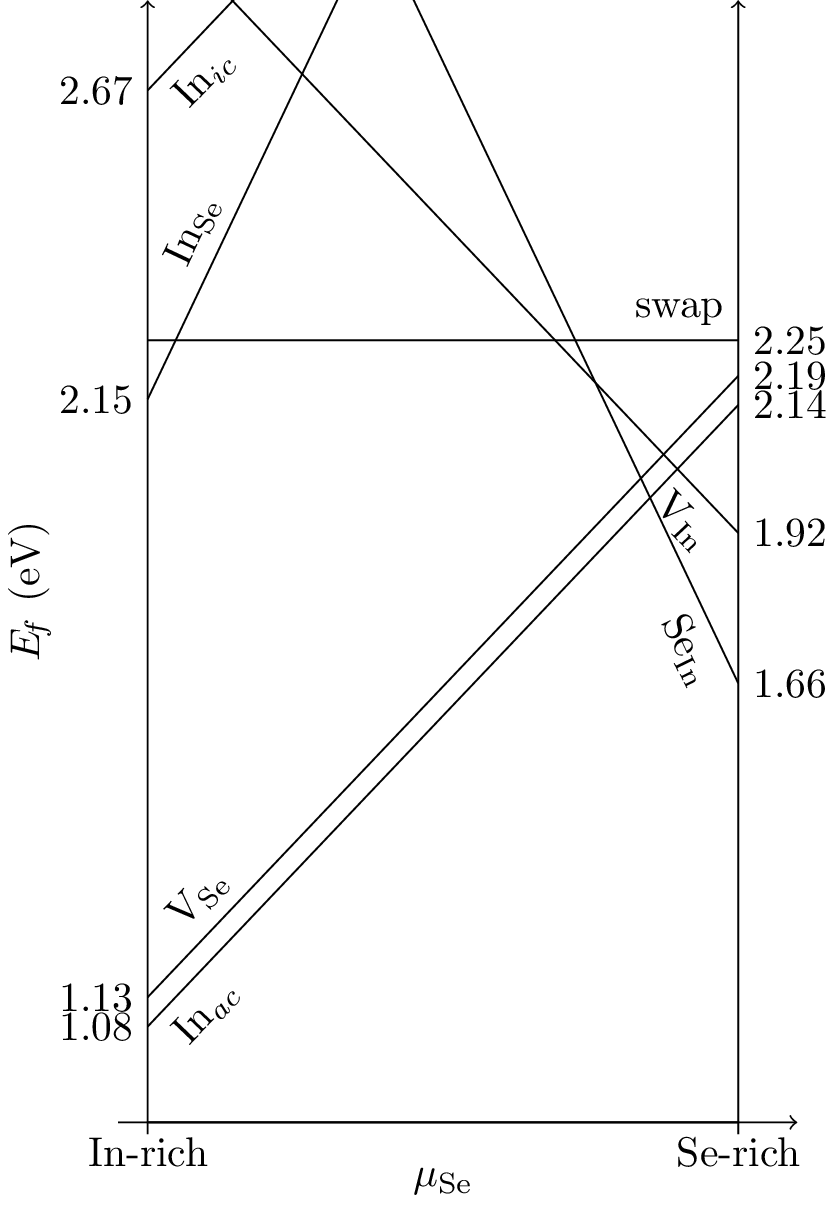}
\caption{\label{fig:chemPot:intrinsic}Formation Energies \( E_{\!f} \)
  as a function of chemical potential \( \mu_{\ch{Se}} \) (arbitrary
  units) for intrinsic defects. \( \Delta \mu_{\ch{Se}} =
  \SI{1.05}{\electronvolt} \). Refer to text for constraints and
  definitions.}
\end{figure}

This work considered seven intrinsic point defects
(Fig.~\ref{fig:ptDxcrysden}): the indium vacancy (V\(_{\ch{In}}\)),
the anti-site defect consisting of a selenium replacing for indium
(\ch{Se}\(_{\ch{In}}\)), indium replacing for selenium
(\ch{In}\(_{\ch{Se}}\)), a swapped In-Se next-neighbor pair
(\ch{In}\(_{\ch{Se}}\)-\ch{Se}\(_{\ch{In}}\)), that we will name
``swap'', the selenium vacancy V\(_{\ch{Se}}\), selenium interstitial
at the hexagonal interstitial site (\ch{In}\(_{ic}\)), and above the
center of the hexagonal interstitial cage (\ch{In}\(_{ac}\)).

The respective band structures are represented in
Fig.~\ref{fig:bs:intrinsic}.  The indium vacancy is a shallow acceptor
(Fig.~\ref{fig:bs:intrinsic}a). \ch{Se}\(_{\ch{In}}\) has a similar
band structure, but the states originating in the In vacancy are
half-filled and move towards mid-gap, whereas the conduction band is
little perturbed (Fig.~\ref{fig:bs:intrinsic}b).  The other anti-site
defect also has semi-filled states, whereas the combined swap of
neighboring \ch{In} and \ch{Se} results in filled defect states near
the valence band (Fig.~\ref{fig:bs:intrinsic}c,d).  The selenium
vacancy introduces defect states both near the valence and conduction
band (Fig.~\ref{fig:bs:intrinsic}e).  Finally, the indium
interstitials are shallow donors (Fig.~\ref{fig:bs:intrinsic}f,g).
The \ch{In}\(_{ac}\) configuration, the most stable (about
\SI{1.59}{\electronvolt} lower in energy than the \ch{In}\(_{ic}\)
configuration), changes little the conduction band dispersion, however
donates free holes to the conduction band states.

The formation energies as a function of the \ch{Se} chemical potential
over all available range are shown in
Fig.~\ref{fig:chemPot:intrinsic}.  As expected, in the \ch{In}-rich
regime the dominant defects are the \ch{Se} vacancy and the \ch{In}
interstitial, whereas in the \ch{Se}-rich limit the dominant defects
are the \ch{In} vacancy and the anti-site where \ch{Se} replaces
\ch{In}.  These regimes will be considered in more detail in the next
sections.

\subsubsection{\ch{In}-rich regime}
\begin{table}
  \caption{\label{tab:dopants} Ionization potential and electron
    affinity \emph{differences} of the various defects in monolayer
    \ch{InSe}, which can be subtracted from \( E_g ( \approx
    \SI{2.4}{\electronvolt}
    )\cite{nanoResearchInSe2014,robertsonUsefulInSe,debbichi2015,olguin2013}
    \) to provide the activation energies via marker method (see
    text). All energies are in \si{\electronvolt}. }
\begin{tabular}{c @{\hskip .5cm} 
  S[ table-format = 1.2 , table-auto-round ] @{\hskip .5cm}
  S[ table-format = 1.2 , table-auto-round ]}
  \hline \hline
  {Defect}  & {\( E_D ( 0 / + ) - E_v \)} & 
    {\( E_c - E_D ( - / 0 ) \)} \\
  \hline
  \(\ch{In}_{ac}\)     & 2.17 &      \\
  V\(_{\ch{Se}}\)       &  .40 &  .65 \\ 
  \(\ch{Se}_{\ch{In}}\) &  .97 & 1.22 \\
  V\(_{\ch{In}}\)       &      & 1.60 \\
  \ch{O2}--A           &      &  .16 \\
  \hline \hline
\end{tabular}
\end{table}

\ch{InSe} crystals are typically grown using the Bridgmann method,
from non-stoichiometric melts with \ch{In} excess, resulting in
\ch{In}-rich crystals.\cite{segura1984electron,
  segura1983investigation, martinez1992shallow}. This is expected due
to the higher volatility of \ch{Se} compared to \ch{In}.

In this regime, the most stable defect, of the four defects we have
considered, is an \ch{In} interstitial above the hexagonal cage,
closely followed by the \ch{Se} vacancy, the latter of which seems to
make a triangular bond between the three \ch{In} atoms surrounding the
vacancy. Both are donors (Fig.~\ref{fig:bs:intrinsic}), with
transition levels at \SI{2.17}{\electronvolt} and
\SI{.4}{\electronvolt} above the valence band, respectively
(Table~\ref{tab:dopants}). In particular, the \ch{In} interstitial,
being a shallow donor, is likely to be the source of the \( n \)-type
conduction in this material, as previously suggested following Hall
effect measurements and position lifetime
experiments\cite{martinez1992shallow, positron-lifetime-InSe,
  segura1983investigation}.  Experimentally, the defect ionization
energy is \SI{18}{\milli\electronvolt}, consistent with the
calculations, that effectively place the transition level close to the
conduction band bottom, within the method
accuracy.\cite{positron-lifetime-InSe} Furthermore, the experimentally
observed donor center concentration is known to increase upon
annealing at \SI{470}{\kelvin} and the donor defects do not affect the
positron lifetime, showing that it is an intrinsic defect and unlikely
to be of vacancy type.\cite{positron-lifetime-InSe} Focusing on the
annealing, we performed a nudged elastic band calculation for both the
indium interstitial and the selenium vacancy in the monolayer case,
obtaining migration activation energies of about
\SI{.21}{\electronvolt} for \ch{In}\(_{ac}\) and
\SI{1.5}{\electronvolt} for V\(_{\ch{Se}}\), in agreement with
expectations. In addition, we note that the anti-site is energetically
expensive, such that it should be rare, and does not contribute to
doping. These establish that the \ch{In} interstitial is responsible
for the \( n \)-type character of undoped samples.

\subsubsection{\ch{Se}-rich regime}
The two relevant intrinsic defects in this regime are the \ch{In}
vacancy and \ch{Se}-replacing-\ch{In} anti-site. V\(_{\ch{In}}\) is a
shallow acceptor, with transition levels calculated to lie
\SI{1.60}{\electronvolt} below the conduction band
(Table~\ref{tab:dopants}). However, since \ch{In} is placed in the
inside of the layer, it is unlikely that V\(_{\ch{In}}\) would exist
on its own, without the removal of neighboring \ch{Se} as well.  \(
\ch{Se}_{\ch{In}} \) is both a donor and an acceptor, with possibly a
negative-\( U \) level ordering (Table~\ref{tab:dopants}).

\subsection{\ch{O2} Physisorption}
\begin{figure*}
\includegraphics[resolution=300,width=\textwidth]{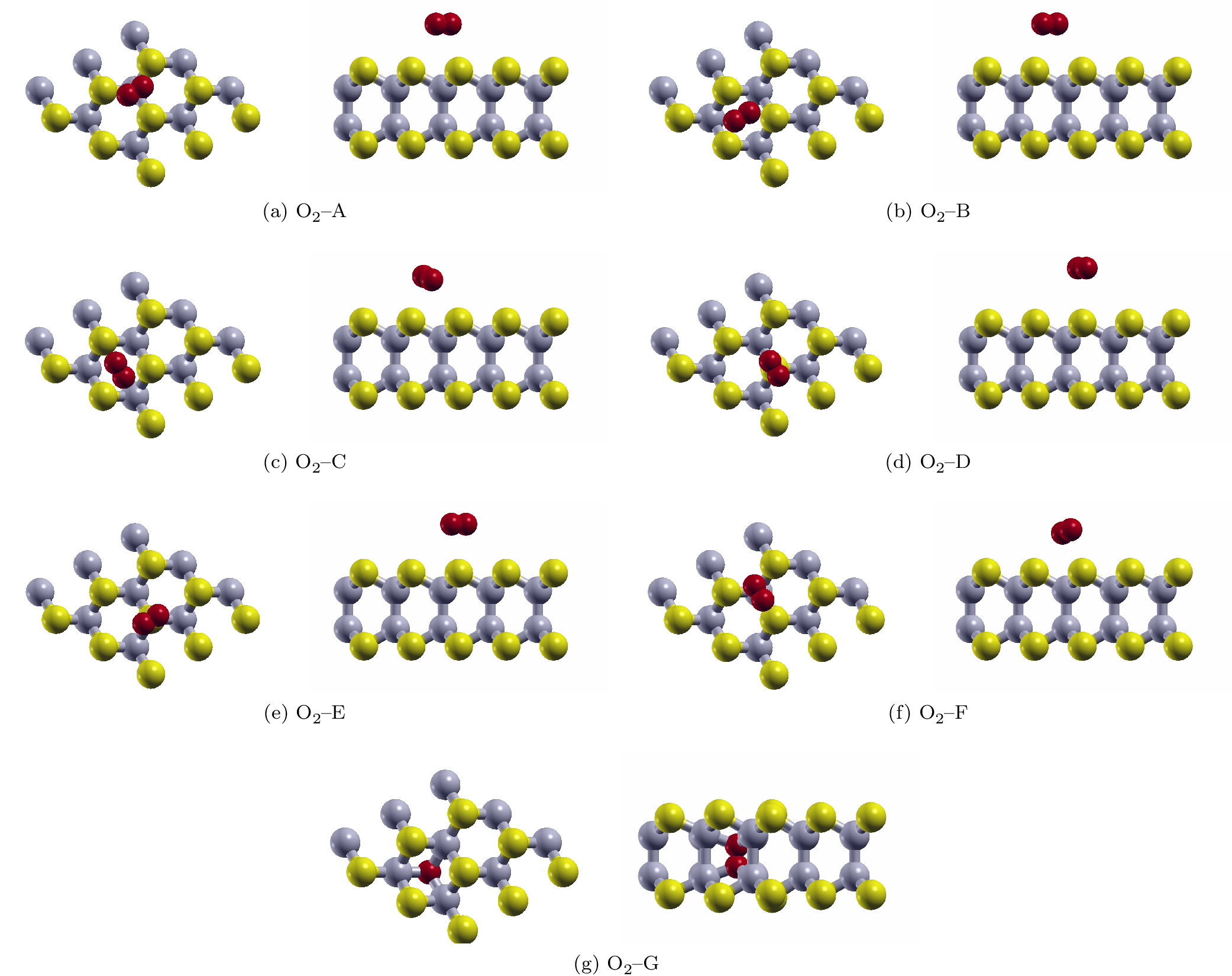}
\caption{\label{fig:addO2xcrysden}(Color online) Top (0001) and side
  (11\(\bar{2}\)0) views of the stable single oxygen molecule
  addition defects in monolayer \ch{InSe} (physisorption), in
  increasing order of relative energy cost of formation. 
  (a) \ch{O2}--A: above indium, perpendicular to bridge bond. 
  (b) \ch{O2}--B: above center of hexagonal cage, perpendicular to
  bridge bonds. 
  (c) \ch{O2}--C: above center of hexagonal cage, along bridge bonds. 
  (d) \ch{O2}--D: above selenium, along bridge bond. 
  (e) \ch{O2}--E: above selenium, perpendicular to bridge bond. 
  (f) \ch{O2}--F: above indium, along bridge bond. 
  (g) \ch{O2}--G: interstitial molecule at center of hexagonal cage,
  perpendicular to monolayer.}
\end{figure*}

\begin{table}
  \caption{Formation energies for each of the various stable oxygen
    absorption defects in monolayer \ch{InSe}. 
    Refer to Fig.~\ref{fig:addO2xcrysden} and Fig.~\ref{fig:addOxcrysden} for
    meaning of abbreviated names. 
    All energies are in \si{\electronvolt}.   
}
\addtocounter{table}{-1} 
\hfill
\subfloat[\label{tab:Ef:addO2}Physisorbed oxygen molecules.]{
\begin{tabular}{c @{\hskip .5cm} S[ table-format = 1.2 , table-auto-round ]}
  \hline \hline
  {Defect}   & { \( E_{\!f} \)  } \\
  \hline
  \ch{O2}--A & -.0162150297 \\
  \ch{O2}--B & -.0158097053 \\ 
  \ch{O2}--C & -.0115210276 \\ 
  \ch{O2}--D & -.0109381542 \\ 
  \ch{O2}--E & -.0035151983 \\ 
  \ch{O2}--F & -.0012833205 \\ 
  \ch{O2}--G &  .9490691946 \\ 
  \hline \hline
\end{tabular}
}
\hfill \hfill 
\subfloat[\label{tab:Ef:addO}Chemisorbed oxygen atoms.]{
\begin{tabular}{c @{\hskip .5cm} S[ table-format = 1.2 , table-auto-round ]}
  \hline \hline
  {Defect}  & { \( E_{\!f} \) } \\
  \hline
  \ch{O}--A & -1.646926308 \\
  \ch{O}--B & -1.637772552 \\ 
  \ch{O}--C &   .048381663 \\ 
  \ch{O}--D &   .368839000 \\ 
  \ch{O}--E &   .741668050 \\ 
  \ch{O}--F &  1.071440388 \\ 
  \ch{O}--G &  2.608849276 \\ 
  \hline \hline
\end{tabular}
}
\hfill
\end{table}

\begin{figure}
\includegraphics[resolution=300,width=\columnwidth]{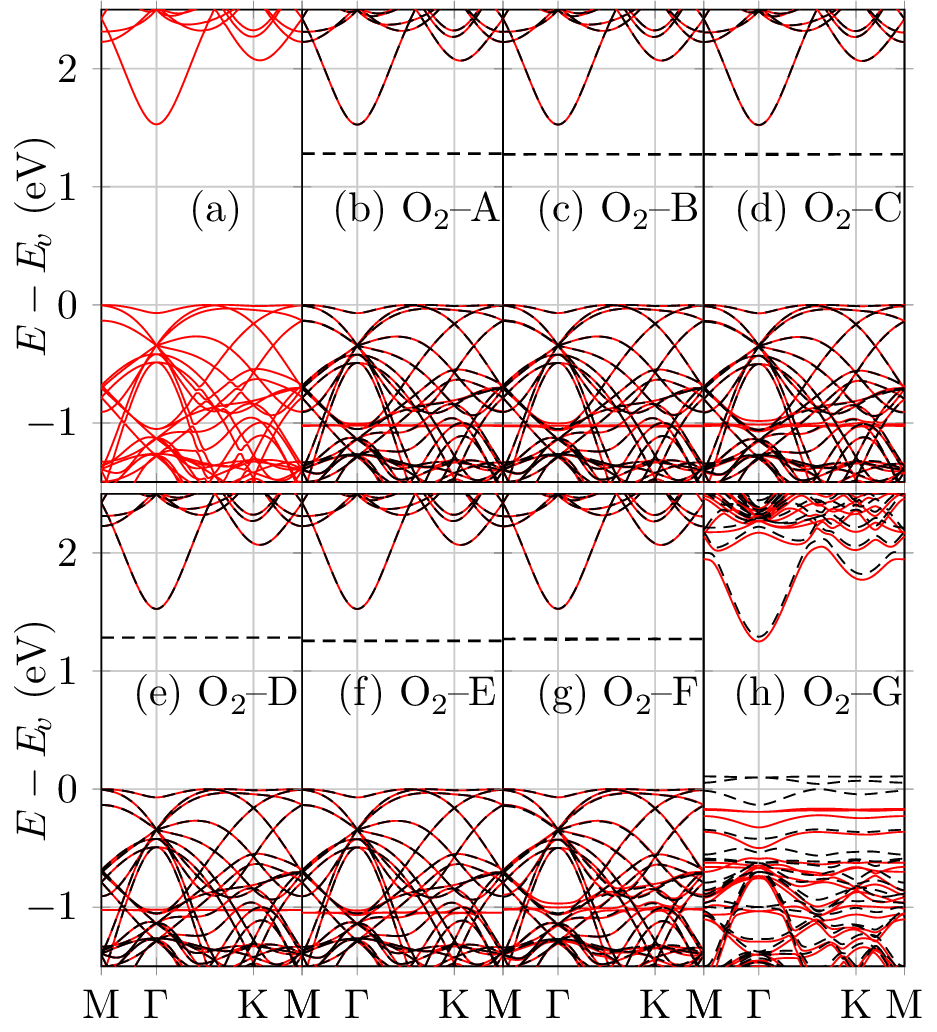}
\caption{\label{fig:bs:addO2}(Color online) DFT band structure plots
  of various stable single oxygen molecule defects in monolayer
  \ch{InSe} (Physisorption) in increasing order of relative energy
  cost of formation. (a) Pristine 3x3 supercell; (b)--(h) different
  configurations of oxygen defects. Refer to
  Fig.~\ref{fig:addO2xcrysden} for the respective defects.  Minority
  spin is shown in dashed line. Color makes deeply embedded impurity
  states easier to see.}
\end{figure}

Figure~\ref{fig:addO2xcrysden} shows the top and side views of all the
possible configurations for oxygen molecule physisorption onto
\ch{InSe}. The formation energies are nearly the same (within
\SI{10}{\milli\electronvolt}) for all the configurations A--F
(Table~\ref{tab:Ef:addO2}). The respective band structures, shown in
Fig.~\ref{fig:bs:addO2}, are also nearly identical, having no gap
states for the majority spin and a double-degenerate empty gap state
for minority spin. The coloring of the band structure plot helps
reveal the deeply embedded impurity states beneath the valence band,
which are flat, similar to the degenerate impurity gap states (dashed
lines) in the band gap. The last of the structures considered,
\ch{O2}--G, consists of an oxygen molecule inside the interstitial
cage. This is \SI{.97}{\electronvolt} higher in energy than surface
physisorbed molecules (Table~\ref{tab:Ef:addO2}). Physisorbed oxygen
can therefore in principle act as electron acceptor, as found in
graphene,\cite{gianozziGrapheneoxygenPhysisorbAcceptor}
phosphorene\cite{cheng-han-2d-4-021007}, and transition metal
dichalcogenides\cite{kumar-PRL}

\subsection{\ch{O} Chemisorption}
\begin{figure*}
\includegraphics[resolution=300,width=\textwidth]{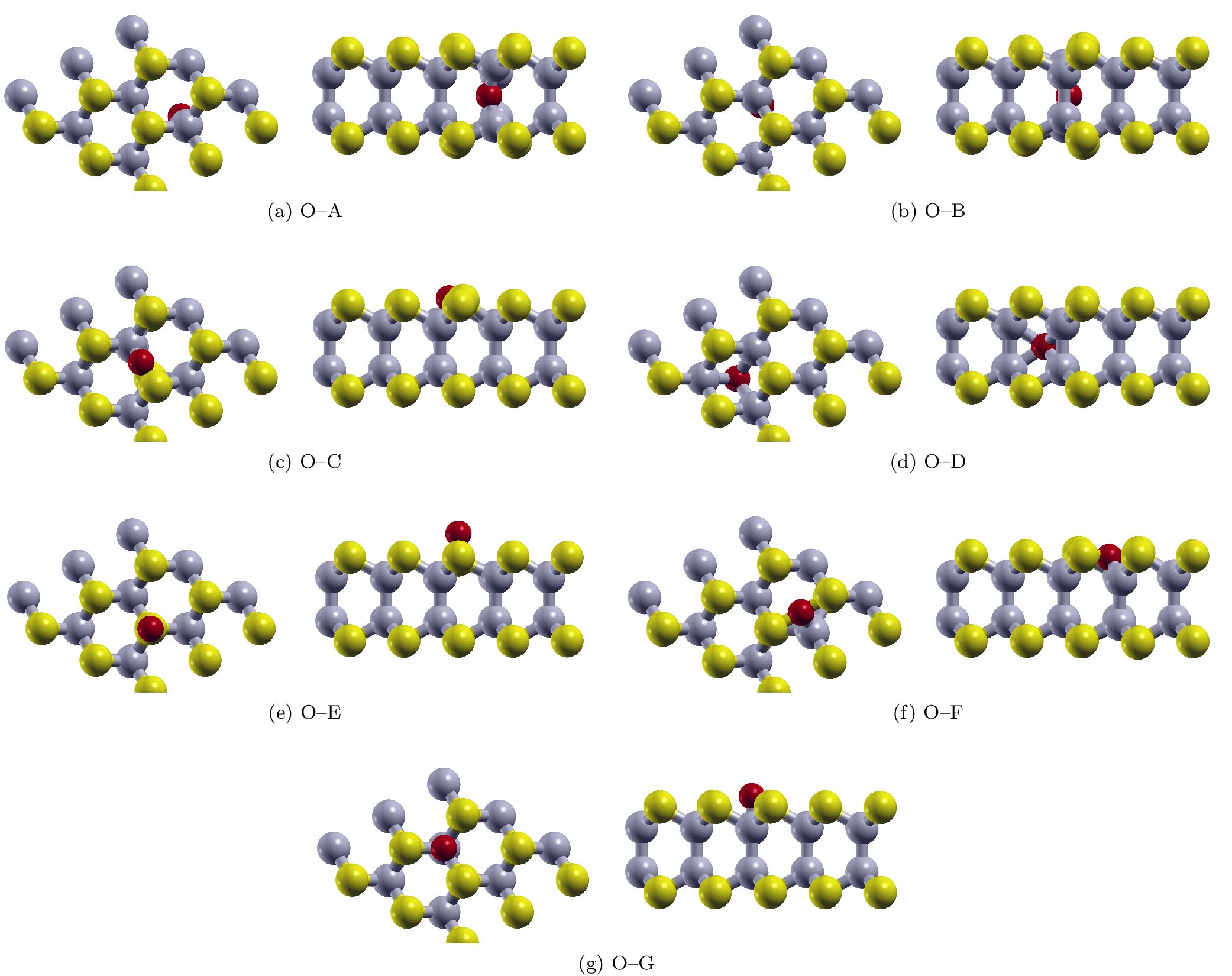}
\caption{\label{fig:addOxcrysden}(Color online) Top (0001) and side
  (11\(\bar{2}\)0) views of the stable single oxygen atom addition
  defects in monolayer \ch{InSe} (Chemisorption), in increasing order
  of relative energy cost of formation. 
  (a) \ch{O}--A: interstitial oxygen defect between two indium atoms,
  with angled bonds like in water molecule, 
  venturing out into the hexagonal interstitial cage. 
  (b) \ch{O}--B: interstitial oxygen in angled bond between two indium
  atoms, underneath (bridge) bond of indium-selenium. 
  (c) \ch{O}--C: oxygen in angled bond between indium and selenium. 
  (d) \ch{O}--D: interstitial oxygen at center of hexagonal cage. 
  (e) \ch{O}--E: oxygen above selenium. 
  (f) \ch{O}--F: three-coordinated oxygen between two selenium atoms,
  also bonded with indium atom. 
  (g) \ch{O}--G: oxygen above indium. 
  The case of oxygen atom hovering above the center of the hexagonal
  interstitial cage is not stable.}
\end{figure*}

\begin{figure}
\includegraphics[resolution=300,width=\columnwidth]{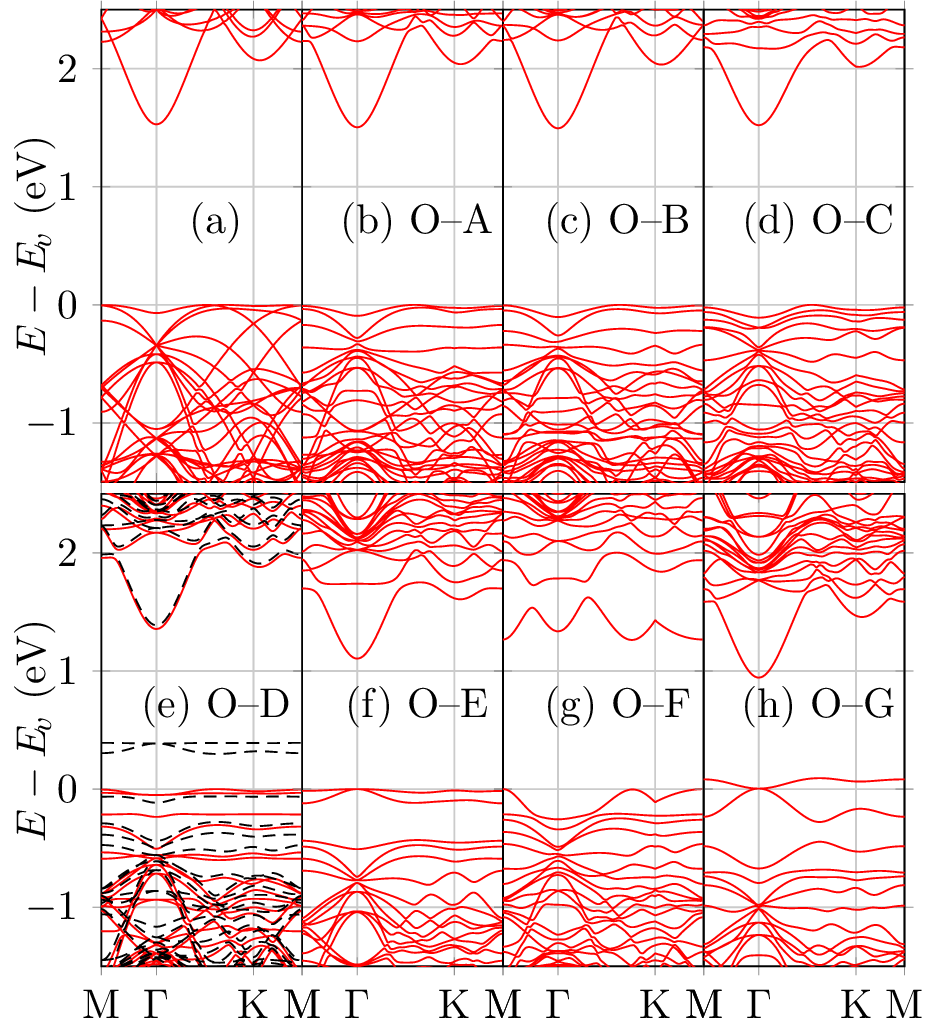}
\caption{\label{fig:bs:addO}(Color online) DFT band structure plots of
  various stable single oxygen atom defects in monolayer \ch{InSe}
  (Chemisorption) in increasing order of relative energy cost of
  formation. (a) Pristine 3x3 supercell; (b)--(h) different
  configurations of oxygen defects. Refer to
  Fig.~\ref{fig:addOxcrysden} for the respective defects. (e) is a
  magnetic spin calculation without spin-orbit coupling. Minority spin
  in dashes.}
\end{figure}

Chemisorption requires breaking the \ch{O2} bond, which is found to
have an energy of \SI{6.61}{\electronvolt} in our calculations, a
typical overestimation, on the high side, under the PBE
approximation\cite{HSEsol} (experimentally measured to be
\SI{5.12}{\electronvolt}\cite{HSEsol}).  Nevertheless, we found that
the chemisorption of oxygen is energetically favorable compared to
physisorption.

Figure~\ref{fig:addOxcrysden} shows the top and side views of all the
single oxygen atom addition defects, while the band structure plots
are presented in Fig.~\ref{fig:bs:addO}. The formation energies \(
E_{\!f} \) do not depend on the \ch{In} and \ch{Se} chemical potentials
(Table~\ref{tab:Ef:addO}).

Table~\ref{tab:Ef:addO} shows that that there is a pair of essentially
degenerate defects that are the lowest in energy.  They are the
\ch{O}--A configuration, interstitial oxygen defect between two indium
atoms, near the bond-center, venturing out into the hexagonal
interstitial cage, and the \ch{O}--B configuration, interstitial
oxygen also near the bond-center between two indium atoms, but
underneath the indium-selenium bond. The other defects are
considerably higher in energy. The band structure plots then tell us
that the three defects of this class, the lowest in energy, are
basically of the same type, and that they barely differ from the band
structure of the PS.

Since chemisorbed oxygen defects have no levels in the gap, their
interaction with vacancies to form substitutional defects will not be
of the Coulomb type but possible strain mediated, since interstitial
atoms, contrary to vacancies, introduce compressing strain on the
surrounding lattice.  In the next section, we will consider the
defects resulting of the interaction between chemisorbed oxygen and
selenium vacancies.

\subsection{\ch{O} Substitution Defects}\label{sec:subO}
\begin{figure}
\includegraphics[resolution=300,width=4.5cm]{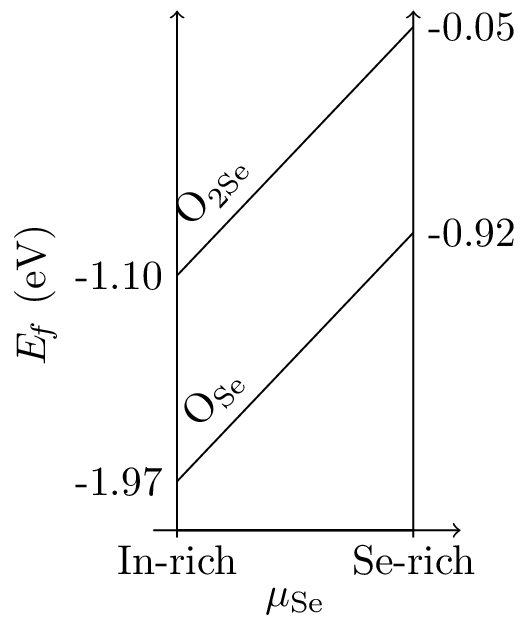}
\caption{\label{fig:chemPot:subO}Formation Energies \( E_{\!f} \) as a
  function of chemical potential \( \mu_{\ch{Se}} \) (arbitrary units)
  for oxygen substitution defects. \( \Delta \mu_{\ch{Se}} =
  \SI{1.05}{\electronvolt} \). Refer to text for constraints and
  definitions.}
\end{figure}

We have considered the possibility that a \ch{Se} lattice site is
occupied by an oxygen atom or by an oxygen molecule
(Fig.~\ref{fig:ptDxcrysden}i,j). The respective band structures are
shown in Fig.~\ref{fig:bs:intrinsic}h,i. The formation energies of
these defects are negative for all the range of chemical potentials,
but are lowest in \ch{In}-rich conditions
(Fig.~\ref{fig:chemPot:subO}). They seem to neither be donors nor
acceptors, just passivating the \( p \)-type selenium vacancy and
reducing the band gap energy. The single substitutional oxygen atom is
\SI{.87}{\electronvolt} lower in energy than the substitutional oxygen
molecule, and it is the most energetically favorable defect presented
in this paper.  It is especially likely to form in the presence of
chalcogen vacancies,\cite{airPassivationChalcogen2D} through the
reaction
\begin{align}
  \frac{1}{2} \ch{O2} + V_{\ch{Se}} \rightarrow \ch{O}_{\ch{Se}}
\end{align}
which has an enthalpy balance of \SI{3.10}{\electronvolt} per oxygen
atom.

\section{Conclusion}
We have investigated the fundamental intrinsic defects in \ch{InSe},
finding that in \ch{Se}-rich material the \( \ch{Se}_{\ch{In}} \)
anti-site is the dominant effect, whereas in \ch{In}-rich material the
indium interstitial and selenium vacancy are the dominant defects. Our
calculations suggest that the unintentional \( n \)-type doping in
cleanly-grown \ch{InSe} should be due to the indium interstitial,
which is a shallow donor, in agreement with arguments from
experiments.

Selenium vacancies have donor deep states at about
\SI{.4}{\electronvolt} above the valence band, that can partially
compensate the doping by interstitials, but this state can be removed
by reaction with molecular oxygen to form substitutional oxygen at the
\ch{Se} site, which has a positive energy balance of
\SI{3.10}{\electronvolt}.

In the absence of intrinsic defects, oxygen chemisorption and
substitution is still energetically favorable, with such defects
having formation energies \( E_{\!f} \) between \num{-.9} and
\SI{-2}{\electronvolt}. Thus, \ch{InSe} monolayers are prone to
oxidation, but still considerably stronger in resilience against the
chemisorption of oxygen than that in phosphorene (the respective
enthalpies for oxygen chemisorption are \SI{-1.65}{\electronvolt} in
\ch{InSe} and \SI{-2.08}{\electronvolt} in
phosphorene\cite{zilettioxygenDefectsPhosphorene}).

We find that chemisorbed oxygen and substitutional oxygen do not have,
in their most stable forms, any ionization levels in the gap. However,
since chemisorbed oxygen atoms are most stable inside the layer and
between In sub-layers, the structural distortion and perturbation of
the charge density distribution induced by chemisorbed oxygen defects
may reduce the carrier mobility, justifying the use of encapsulating
layers in \ch{InSe}-based electronic devices.

\section*{Acknowledgements}
This work was supported by the National Research Foundation, Prime
Minister Office, Singapore, under its Medium Sized Centre Programme
and CRP award ``Novel 2D materials with tailored properties: beyond
graphene" (Grant number R-144-000-295-281).  The first-principles
calculations were carried out on the CA2DM high-performance computing
facilities.

%
\end{document}